\documentclass[preprint,showpacs,preprintnumbers,amsmath,amssymb,superscriptaddress]{revtex4}

\usepackage{graphicx,amsfonts}
\usepackage{epsfig}
\usepackage{dcolumn}
\usepackage{bm}

\begin{document}

\title{Minimal 3-3-1 model with a spectator sextet }


\author{G. De Conto}%
\email{georgedc@ift.unesp.br}
\affiliation{
Instituto  de F\'\i sica Te\'orica--Universidade Estadual Paulista \\
R. Dr. Bento Teobaldo Ferraz 271, Barra Funda\\ S\~ao Paulo - SP, 01140-070,
Brazil
}
\author{A. C. B. Machado}%
\email{ana@ift.unesp.br}
\affiliation{
Instituto  de F\'\i sica Te\'orica--Universidade Estadual Paulista \\
R. Dr. Bento Teobaldo Ferraz 271, Barra Funda\\ S\~ao Paulo - SP, 01140-070,
Brazil
}
\author{V. Pleitez}%
\email{vicente@ift.unesp.br}
\affiliation{
Instituto  de F\'\i sica Te\'orica--Universidade Estadual Paulista \\
R. Dr. Bento Teobaldo Ferraz 271, Barra Funda\\ S\~ao Paulo - SP, 01140-070,
Brazil
}

\date{10/20/15}
%
\begin{abstract}
We consider the minimal 3-3-1 model with a heavy scalar sextet and realize, at the tree level, an effective dimension-five interaction that contributes to the mass of the charged leptons. In this case the charged leptons masses arise from a sort of type-II seesawlike mechanism while the neutrino masses are generated by a type-I mechanism. We also show that the parameters giving the correct lepton masses also accommodate the Pontecorvo-Maki-Nakawaga-Sakata matrix.  We  give the scalar mass spectra of the model and analyze under which conditions the fields in the scalar sextet are heavy even with small or zero vacuum expectation values.
We also show the conditions under which it is possible to have a stable (bounded from below) potential and also a global minimum.
\end{abstract}


\maketitle

\section{Introduction}
\label{sec:intro}

Although we have known since 2012 that there exists a neutral spin-0 resonance with properties (mass and couplings) that are compatible within the experimental error, with those of the scalar SM-Higgs boson~ \cite{Aad:2012tfa, Chatrchyan:2012xdj}, these data do not exclude the existence of more fields of this sort.
In fact, almost all the extensions of the standard model (SM) include extra scalar 
multiplets: complex~\cite{Chikashige:1980ui} or real singlets~\cite{Hill:1987ea,Davoudiasl:2004be,vanderBij:2006ne}, two~\cite{Branco:2011iw} or more doublets~\cite{Machado:2012ed}, and Hermitian~\cite{Brdar:2013iea} and/or non-Hermitian triplets~\cite{Konetschny:1977bn,Magg:1980ut,Cheng:1980qt,Escobar:1982dp}. Moreover, extra scalar multiplets usually are introduced in a given model just in order to give masses to the neutrino and/or charged leptons. Hence, they may have small vacuum expectation values (VEV). 
However, this usually implies that there might be light neutral scalars which can be easily ruled out by phenomenology. In two-Higgs doublets this is not the case when there is a positive quadratic term $\mu^2>0$, which behaves like a positive mass square term in the scalar potential. In this case such parameters may dominate the contributions to the masses of the multiplets' members, which are almost mass degenerated, i.e. in the context of models with one or more inert doublets, see~\cite{Machado:2012ed} and \cite{Ma:2006km}. 

The 3-3-1 models are intrinsically multi-Higgs models. For instance,
in the minimal 3-3-1 model (here denoted by m331 for short), the charged leptons gain mass from a triplet and a sextet and the neutrino gain Majorana masses only through  the sextet ~\cite{Pisano:1991ee,Foot:1992rh,Frampton:1992wt}. 
On the other hand, in the model with heavy charged~\cite{Pleitez:1992xh} or neutral leptons~\cite{Montero:1992jk}, only the triplets are needed, if right-handed neutrinos are introduced and the type-I seesaw mechanism is implemented. 

Because of the sextet, the scalar potential in the m331 becomes more complicated, for this reason it was pointed out in Ref.~\cite{Montero:2001tq}  that the sextet can be omitted if a dimension-five effective operator, involving only triplets, is in charge of the mass generation of the charged leptons and neutrinos.  The m331 model without the sextet was called the "reduced" m331 model in Ref.~\cite{Ferreira:2011hm} because only the triplets $\rho$ and $\chi$ are introduced. However, there are important differences between our model and those in Refs.~\cite{Montero:2001tq,Ferreira:2011hm} as we will discuss in Sec.~\ref{sec:con}. Moreover, as in the case of the SM, the question now is, how does this effective operator arise at tree and/or loop level~\cite{Ma:1998dn,Bonnet:2012kz,Sierra:2014rxa}? In the context of the m331 model, mechanisms for generating effective dimension-five operators for the case of the neutrino masses were given in Ref.~\cite{Montero:2001ts}. 
However, in those works thePontecorvo-Maki-Nakawaga-Sakata (PMNS) matrix was not considered. It is far from obvious that the same parameters that allow to obtain the correct lepton masses also accommodate a realistic PMNS matrix. We show that this is possible in the m331 model with a heavy sextet which implements a sort of type-II seesawlike mechanism in the charged lepton sector and, by introducing right-handed neutrinos, neutrino masses arise from a type-I seesaw mechanism. 

In fact, we show that the sextet is just a way to generate, at the tree level, the effective five-dimensional operator proposed in Ref.~\cite{Montero:2001tq} in order to give the charged leptons their correct masses. This happens if all fields in this multiplet are heavy and its neutral components gain a small ($s^0_2$) or a zero ($s^0_1$) VEV. We also study in this model the conditions upon the dimensionless coupling constants that imply a scalar potential bounded from below, with a global minimum as well. We obtain a realistic PMNS mixing matrix as well. 

The outline of this paper is the following. In Sec.~\ref{sec:scalars} we give the scalar representation content of the m331 model and the scalar potential of the model. In Sec.~\ref{sec:masses} we obtain the scalar mass spectra of the model under the conditions of $Z_7\otimes Z_3$ discrete symmetries. In Sec.~\ref{sec:pmns} we obtain the charged lepton and neutrino masses and the corresponding PMNS matrix.
Our conclusions appear in Sec.~\ref{sec:con}. The conditions for a stable minimum, at tree level, of the scalar potential are obtained in Appendix~\ref{sec:appendixa}. In Appendix B we consider the Goldstone bosons in the model with exact mass matrices. 

\section{The scalar sector in the m331 model}
\label{sec:scalars}

The scalar potential in several 3-3-1 models was considered in Refs.~\cite{Diaz:2003dk,Nguyen:1998ui,Giraldo:2011gd,Hernandez:2014vta}. Here, however, we
will study only the  m331 model in the situation in which the sextet gain a small VEV, the extra scalars in the sextet are heavy,  and there is no explicit total lepton number violation in the scalar potential, which is avoided by an appropriate discrete symmetry.

In the m331 model the scalar sector is composed of a sextet $S\sim(\textbf{6},0)$ 
and three  triplets: $\eta=(\eta^0\,\;-\!\eta^{-}_1\,\eta^+_2)^T\sim({\bf3},0)$,
$\rho=(\rho^+\,\rho^0\,\rho^{++})^T\sim({\bf3},1)$, and
$\chi=(\chi^-\,\chi^{--}\,\chi^0)^T\sim({\bf3},-1)$, where $(x,y)$ refer to the $(SU(3)_L,U(1)_X)$ transformations. Only the triplet $\eta$ and the sextet~$S$,
\begin{equation}
S=\left(
\begin{array}{ccc}
s^0_1& \frac{s^-_1}{\sqrt2} & \frac{s^+_2}{\sqrt2}\\
\frac{s^-_1}{\sqrt2}& S^{--}_1&\frac{s^0_2}{\sqrt2}\\
\frac{s^+_2}{\sqrt2}&\frac{s^0_2}{\sqrt2}&S^{++}_2
\end{array}
\right),
\label{sextet1}
\end{equation}
couple to the leptons through the Yukawa interactions $\overline{(\Psi_L)^c}\Psi_LS^*$ and $\overline{(\Psi_L)^c}\Psi_L\eta$. 

We can write the $SU(3)$ multiplets above in terms of the $SU(2)$ ones. For the triplets we write
\begin{equation}
\eta=
\left(
\begin{array}{c}
\Phi_\eta \\ \eta^+_2\end{array}\right)\sim({\bf3},0),\quad
\rho=\left(\begin{array}{c}
\Phi_\rho \\ \rho^{++}\end{array} \right)\sim({\bf3},1),\quad
\chi=\left(
\begin{array}{c}
\Phi_\chi\\ \chi^0\end{array} \right)\sim({\bf3},-1).
\label{tripletos}
\end{equation}
The sextet in Eq.~(\ref{sextet1}) can be written as
\begin{equation}
S=\left(
\begin{array}{cc}
T & \frac{\Phi_s}{\sqrt2} \\
\frac{\Phi^t_s}{\sqrt2}  & S^{--}_2
\end{array}
\right),\quad S^*=\left(
\begin{array}{cc}
T^* & \frac{\Phi^*_s}{\sqrt2} \\
\frac{\Phi^\dagger_s}{\sqrt2}  & S^{++}_2
\end{array}
\right),
\label{sextet2}
\end{equation}
where $\Phi^t_s$ means the transpose of the doublet $\Phi_s$. Under the $SU(2)\otimes U(1)_Y$ group the multiplets $\Phi_{\eta,\rho,\chi,s}$  in Eqs.~(\ref{tripletos}) and (\ref{sextet2}) transform as
\begin{equation}
\Phi_\eta=\left(
\begin{array}{c}
\eta^0\\ -\eta^-_1
\end{array}
\right),\;  \Phi_\rho=\left(
\begin{array}{c}
\rho^+\\ \rho^0
\end{array}
\right),\; \Phi_\chi=\left(
\begin{array}{c}
\chi^-\\ \chi^{--}
\end{array}
\right), \; \Phi_s=\left(
\begin{array}{c}
s^+_2\\ s^0_2
\end{array}
\right),
\label{dubletos}
\end{equation}
where these are doublets with weak hypercharge $Y=-1,+1,-3,+1$,  and $T$ in Eq.~(\ref{sextet2})
\begin{equation}
T=\left( \begin{array}{cc}
s^0_1 &\frac{s^+_1}{\sqrt2} \\
\frac{s^+_1}{\sqrt2} & S^{--}_1
\end{array}
\right),
\label{triplet}
\end{equation}
is a triplet with $Y=2$. The $SU(2)$ singlets $\eta^+_2,\rho^{++},\chi^0,S^{--}_2$ have $Y=+2,+4,0,+4$, respectively. 

The total lepton number assignment in the scalar sector is~\cite{Liu:1993gy}
\begin{equation} 
L(T^*,\eta^-_2,\Phi_\chi,\rho^{--},S^{--}_2)=+2,\quad L(\Phi_{\eta,\rho,s},\chi^0)=0.
\label{ln}
\end{equation}
Notice that the  only scalar doublet carrying lepton number is $\Phi_\chi$, and both members of the doublet have electric charge; for this reason, $\langle \Phi_\chi\rangle=0$ always. The existence of scalars carrying lepton number implies the possibility of explicit breaking of this quantum number in the scalar potential. It is possible to avoid such terms by imposing an appropriate discrete symmetry. We show one possibility in Table~\ref{z7}.
In the table, $Q_{1,2}$ denote the quark triplets and $Q_3$, the quark antitriplet, $j_{mR}$ and $J$ are exotic quarks carrying electric charge of -4/3 and 5/3, in units of the positron charge $e$. For more details, see Ref.~\cite{Machado:2013jca}.  

\begin{figure}
\begin{center}
\includegraphics[width=5in]{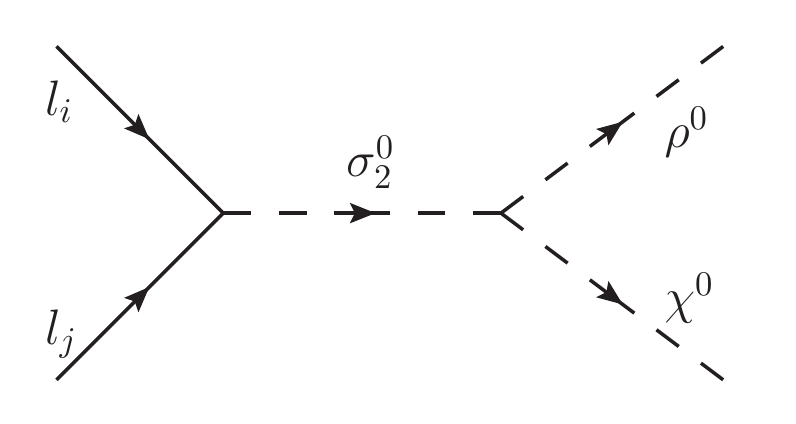}
\caption{Tree-level realization of the effective dimension-five operator in Eq.~(\ref{effective1}) with heavy scalar sextet.}
\label{fig1}
\end{center}
\end{figure}

Since the complex triplet  $T$ and the singlet (under $SU(2)$) $S^{++}_2$   carry lepton number they do not mix with  $\Phi_{\eta,\rho,\chi}$ if there are no lepton number violating terms in the scalar potential. As we will show below, there is some range of the parameter space that allows $\langle s^0_1\rangle=0$ and $\langle s^0_2\rangle/v_W\ll1$, where $v_W=246$ GeV is the electroweak energy scale.  In this situation the neutral scalar $s^0_1$ does not participate in the spontaneous symmetry breaking and $s^0_2$ has a small effect on the vector and charged lepton masses. At this stage, active neutrinos are massless and the charged leptons gain a rather small mass. However, these scalar fields are heavy and the charged leptons gain the appropriate mass through the interaction with the triplet $\eta$ and an effective interaction involving the triplets $\rho$ and $\chi$. Similar to the standard model a non-Hermitian scalar triplet generates, at tree level, the neutrino masses by the interaction $(1/\Lambda)\phi^0\phi^0\nu\nu$ through the exchange of a complex triplet~\cite{Ma:1998dn} (see  Fig.~\ref{fig1}).

\begin{table}
	\centering
	\begin{tabular}{|c|c|c|c|c|c|c|c|c|c|c|c|c|}\hline\hline
		& $Q_{(1,2)L}$ & $Q_{3L}$ & $U_{aR}$ & $D_{aR}$ & $\Psi_{aL}$ & $\nu_{aR}$ & $\eta$ & $\chi$ & $\rho$ & $S$ & $j_{mR}$ & $J_{R}$ \\ \hline
		$Z_{7}$ & 1 & $\omega^6$ & $\omega^5$ & $\omega^3$ & $\omega^2$ & $\omega^6$ & $\omega^3$ & $\omega^6$
		& $\omega^5$ & $\omega^4$ & $\omega^6$ & $\omega^2$ \\ \hline
		$Z_{3}$ & 1 & w$^2$ & w  & w & w  & w  & w  & 1 
		& w & w  & 1 & w \\ \hline \hline
	\end{tabular}
	\caption{Transformation properties of the fermion and scalar fields under $Z_{7} \otimes Z_{3} $. Here $\omega=e^{i2\pi/7}$ and w$=e^{i2\pi/3}$. } 
	\label{z7}
\end{table}

The most general scalar potential involving the three triplets and the sextet is~\cite{Liu:1993gy}
\begin{equation}
V(\eta,\rho,\chi,S)=V^{(2)}+V^{(3)}+V^{(4a)}+\cdots +V^{(4e)},
\label{potential}
\end{equation}
where 
\begin{eqnarray}
V^{(2)}&=&\sum_{X=\eta,\rho,\chi,S}\mu^2_X\textrm{Tr}(X^\dagger X),\nonumber \\
V^{(3)}&=& \frac{1}{3!} \,f_1\epsilon_{ijk} \eta_i\rho_j\chi_k+f_2 (\chi^T S^* \rho+\rho^T S^*\chi) +f_3 \eta^T S^* \eta+
\frac{f_4}{3!}\,\epsilon_{ijk}\epsilon_{mnl}\;S^*_{im}S^*_{jn}S^*_{kl} ,
\nonumber \\
V^{(4a)}&=&a_1(\eta^\dagger \eta)^2+a_2(\rho^\dagger\rho)^2+ a_3(\chi^\dagger \chi)^2+
\chi^\dagger\chi(a_4\eta^\dagger\eta+a_5\rho^\dagger\rho)+a_6(\eta^\dagger\eta)(\rho^\dagger \rho)\nonumber\\&& 
a_7(\chi^\dagger \eta)(\eta^\dagger\chi)+a_8(\chi^\dagger\rho)(\rho^\dagger\chi)+
a_9(\eta^\dagger\rho)(\rho^\dagger\eta)+[a_{10}(\chi^\dagger\eta)(\rho^\dagger\eta)+H.c.],\nonumber \\
V^{(4b)}&=& b_1\chi^\dagger S\hat{\chi}\eta+b_2\rho^\dagger S\hat{\rho}\eta+b_3\eta^\dagger S[\hat{\chi}\rho-\hat{\rho}\chi]+H.c.,\nonumber \\
V^{(4c)}&=&   c_1\textrm{Tr}[\hat{\eta}S\hat{\eta}S]+c_2\textrm{Tr}[\hat{\rho}S\hat{\chi}S]+H.c.,\nonumber \\
V^{(4d)}&=&d_1(\chi^\dagger \chi)\textrm{Tr}S S^*+d_2 [(\chi^\dagger S)(S^* \chi)]+d_3(\eta^\dagger\eta)\textrm{Tr}(S S^*)+d_4\textrm{Tr}[(S^*\eta)(\eta^\dagger S)]\nonumber \\
&&+d_5(\rho^\dagger\rho)\textrm{Tr}S S^*+d_6 \textrm{Tr}[(S^* \rho)(\rho^\dagger S)],\nonumber \\
V^{(4e)}&=& e_1(\textrm{Tr}S S^*)^2+e_2\textrm{Tr}(S S^*S S^*),
\label{potential2}
\end{eqnarray} 
and we have defined in the $V^{(b)}$ and $V^{(c)}$ terms $\hat{x}_{ij}= \epsilon_{ijk}x_k$, with $x=\eta,\rho,\chi$. Notice also that $S^\dagger=S^*$ since $S$ is a symmetric matrix. The conditions for having a potential bounded from below in Eq.~(\ref{potential2}), under the conditions in Table~\ref{z7}, are given in Appendix~\ref{sec:appendixa}.

Concerning the vacuum alignment and the conservation of the lepton number $L$, five possibilities can be considered (see also Ref.~\cite{Liu:1993gy}) 
\begin{enumerate}
\item[a)] Explicit $L$ violation and $\langle s^0_{1,2}\rangle\not=0$ and arbitrary. This is the most general case and it has not been consider in the literature.  
\item[b)] Explicit lepton number violation in the scalar potential and  $\langle s^0_1\rangle = 0$ and  $\langle s^0_2\rangle\not=0$ at tree level, but $\langle s^0_1\rangle\not=0$  by loop corrections  \cite{Pleitez:1992xh,Frampton:1993wu}.
\item[c)] No explicit $L$ violation and $\langle s^0_1\rangle=0$ and $\langle s^0_2\rangle=0$. Notwithstanding, the latter condition is not stable under radiative corrections unless a fine-tuning
is imposed. 
 
\item[d)] No explicit lepton number violation but $\langle s^0_1\rangle\not=0$ and $\langle s^0_2\rangle\not=0$. In this case there is a triplet Majoron that has been ruled out by the $Z$ invisible width~\cite{GonzalezGarcia:1989zh}.

\item[e)] No explicit $L$ violation and $\langle s^ 0_1\rangle=0$ but $\langle  s^0_2\rangle\not=0$. Although $L$ is conserved, there is violation of the family numbers $L_{e,\mu,\tau}$. In this case $\langle s^0_1\rangle=0$ is stable at tree and higher-order level~\cite{Foot:1992rh,Pleitez:1992xh}. 
\end{enumerate}

Here we will consider the last case, (e), with $\langle s^0_1\rangle=0$ and $\langle s^0_2\rangle/v_W \ll 1$.  
This case occurs if the constraint $v^2_W=\sum v^2_i=(246\,\textrm{GeV})^2$ (note that $i = \rho, \eta, s_1, s_2$) is saturated with the $v^2_\eta$ and $v^2_\rho$ as in Ref.~\cite{Machado:2013jca}. 
Moreover, as we said before, in order to simplify the scalar potential, we impose a $Z_7$ discrete symmetry which forbids the $L$ violating terms, $f_3,f_4,a_{10},b_3,c_2=0$, but also the terms $c_1$ and $b_{1,2}$ are forbidden if we impose an additional discrete $Z_3$ symmetry. See Table~\ref{z7}.

\section{Scalar mass spectra in the model}
\label{sec:masses}
 
Let us consider the scalar potential in Eq.~(\ref{potential2}) with the $Z_{7}\otimes Z_3$ symmetries given in Table~\ref{z7}. We make as usual $y^0=(1/\sqrt{2})(v_y+X^0_y+iI^0_y)$, where $y=\eta,\rho,\chi$ and $s_2$.
The constraint equations, obtained by imposing that $\partial V/\partial v_y=0$, being $V$ the potential in (\ref{potential2}), under the conditions of the item e) above and considering all VEVs real, are given by
\begin{eqnarray}
&& v_\eta\left[\mu^2_\eta+a_1v^2_\eta+\frac{a_6}{2}v^2_\rho+\frac{a_4}{2}v^2_\chi+ d_3v^2_{s_2}+\frac{f_1v_\rho v_\chi}{2\sqrt{2}v_\eta}\right]=0,\nonumber \\&&
v_\rho\left[\mu^2_\rho+\frac{a_6}{2}v^2_\eta+ a_2v^2_\rho+\frac{a_5}{2}v^2_\chi+ \frac{d_{56}}{2} v^2_{s_2}+\frac{(f_1v_\eta+f_2v_{s_2})v_\chi} {2\sqrt{2}v_\rho}\right]=0,\nonumber \\&&
v_\chi\left[\mu^2_\chi+\frac{a_4}{2}v^2_\eta+\frac{a_5}{2}v^2_\rho+a_3v^2_\chi+ d_{12}v^2_{s_2}+\frac{(f_1v_\eta v_\rho+f_2v_{s_2})v_\rho}{2\sqrt{2}v_\chi}\right]=0,\nonumber \\ && 
v_{s_2}\left[2\mu^2_S+d_3 v^2_\eta+\frac{d_{56}}{2} v^2_\rho+\frac{d_{12}}{2}v^2_\chi+2e_{12}v^2_{s_2}+\frac{f_2v_\rho v_\chi}{2\sqrt{2}v_{s_2}}\right]=0,
\label{ce}
\end{eqnarray} 
where we have defined $d_{56}=2d_5+d_6$, $d_{12}=2d_1+d_2$, and $e_{12}=2e_1+e_2$.
Notice that no VEV can be zero unless $f_1=f_2=0$. However, if this were the case, the scalar potential has a non-Abelian symmetry larger than the rest of the Lagrangian. Hence, $f_1,f_2\not=0$ in order that the gauge symmetry of the scalar potential is the same as the other terms of the Lagrangian. 
In the case of the sextet, even if we had begun with $\mu^2_S>0$ and $v_{s_2}=0$, the term $f_2\not=0$ induces a tadpole which implies a counterterm leaving this VEV arbitrary. Hence, we assume $\mu_S^2<0$ and $v_{s_2}\not=0$ but is small in the sense that $v_{s_2}/v_W\ll1$. In Appendix~\ref{sec:appendixb} we show explicitly the Goldstone bosons. 

Here, we will consider the mass matrices of the $C\!P$-even scalars and other mass matrices in Appendix~\ref{sec:appendixb}, assuming that $v_{s_2}/v_\chi<<1$. We also disregard some off-diagonal terms besides the ones from the assumption just mentioned, assuming that the respective diagonal elements are much bigger, to further simplify the matrices. In this approximation it is possible to obtain exact eigenvectors, but the case of the $CP$-even neutral scalars is more complicated and we will not consider here in detail. We show only that at least the two neutral scalars in the sextet are heavy. Analytical expressions, within the approximation above, of the masses and the eigenstates of the neutral $C\!P$-odd and the charged sectors are given.
 
\subsection{Neutral $C\!P$-even scalars}
\label{subsec:cpeven2}

In this sector the mass matrix $m^2$ is $5\times5$ decompose, in the approximation used, into $4\times4$ + $1\times1$,  where 
the $4\times4$ matrix, in the basis $(X_\eta^0,X_\rho^0, X_\chi^0, X_{s2}^0 )$, is given by
\begin{equation}
\left(
\begin{array}{cccc}
2 a_1 v_\eta^2-\frac{f_1 v_\rho v_\chi}{\sqrt{2} v_\eta} & 	 a_6v_\eta v_\rho+\frac{f_1 v_\chi}{\sqrt{2}} & \frac{f_1 v_\rho}{\sqrt{2}}+a_4 v_\eta v_\chi & d_3 v_\eta v_{s_2} \\
& \frac{4 a_2 v_\rho^3-(\sqrt{2} f_1 v_\eta -f_2 v_{s_2}) v_\chi}{2 v_\rho} & \frac{f_1 v_\eta}{\sqrt{2}}+\frac{f_2 v_{s_2}}{2}+a_5 v_\rho v_\chi & \frac{1}{2} [(2 d_5+d_6) v_\rho v_{s_2}+f_2 v_\chi] \\
& & \frac{4 a_3 v_\chi^3-\sqrt{2} f_1 v_\eta v_\rho-f_2 v_\rho v_{s_2}}{2 v_\chi} & \frac{1}{2} [f_2 v_\rho+(2 d_1+d_2) v_{s_2} v_\chi] \\
&  &  & (2 e_1+e_2) v_{s_2}^2-\frac{f_2 v_\rho v_\chi}{2 v_{s_2}} \\
\end{array}
\right),
\label{cpeven1}
\end{equation}
and the $1\times1$ part implies the eigenvalue is $m^2_5=\frac{(d_2  v_\chi^2+2 d_4 v_\eta^2 -d_6 v_\rho^2 -2 e_2 v_{s_2}^2)v_{s_2}-2 f_2 v_\rho v_\chi}{4 v_{s_2}}>0$, which implies
$d_2v^2_\chi/4-2 f_2 v_\rho v_\chi/v_{s_2}>(2d_4v^2_\eta- d_6 v_\rho^2 -2 e_2 v_{s_2}^2)/4$. Recall that $f_2<0$.
Here the mass eigenstates are denoted by $m^2_i,\;i=1,\cdots,5$.
In fact, we can see that $m^2_5\approx (1/4)d_2v^2_\chi-f_2v_\rho v_\chi/v_{s_2}$; hence, this is a large mass.  One of them must be mainly that of the 125 GeV discovery at the LHC. Although we have not given details, according to the results in Ref.~\cite{Machado:2013jca}, if $Re\,\rho^0=0.42 h+\cdots$ this scalar has the same coupling with the top quarks that is numerically equal to that from the Higgs boson in the SM.  

\subsection{Neutral CP-odd scalars}
\label{subsec:cpodd2}

The mass matrix in Eq.~(\ref{i2}), in the approximation of subsection ~\ref{subsec:cpeven2},
also decomposes in  $3\times3$ and  diagonal $2\times2$ matrices. The $3\times3$ matrix,
in the basis $(I_\eta^0,I_\rho^0, I_\chi^0) $, is
\begin{equation}
M^2\approx \frac{1}{2} \left(
\begin{array}{ccc}
\frac{f_1 v_\rho v_\chi}{\sqrt{2} v_\eta} & \frac{f_1 v_\chi}{\sqrt{2}} & \frac{f_1 v_\rho}{\sqrt{2}} \\
\frac{f_1 v_\chi}{\sqrt{2}} & \frac{f_1 v_\eta v_\chi}{\sqrt{2} v_\rho} & \frac{f_1 v_\eta}{\sqrt{2}} \\
\frac{f_1 v_\rho}{\sqrt{2}} & \frac{f_1 v_\eta}{\sqrt{2}} & \frac{f_1 v_\eta v_\rho}{\sqrt{2} v_\chi} \\
\end{array}
\right).
\label{i1}
\end{equation}

This matrix has two zero eigenvalues and another nonzero one: 
\begin{equation}
\begin{array}{cl}
M^2_1&=M^2_2=0 \\
M^2_3&= -\frac{f_1 \left(v_\eta^2 \left(v_\rho^2+v_\chi^2\right)+v_\rho^2 v_\chi^2\right)}{\sqrt{2} v_\eta v_\rho v_\chi}\approx -\frac{f_1v_\eta v_\chi}{\sqrt{2}v_\rho}>0,\;\;f_1<0,
\label{a1}
\end{array}
\end{equation}
with respective eigenvectors given by the columns of the matrix:
\begin{equation}
\left(
\begin{array}{ccc}
-\frac{v_\eta}{\sqrt{v_\eta^2+v_\chi^2}} & -\frac{v_\eta v_\chi^2}{\sqrt{\left(v_\eta^2+v_\chi^2\right) \left(\left(v_\rho^2+v_\chi^2\right) v_\eta^2+v_\rho^2 v_\chi^2\right)}} & \frac{v_\chi}{v_\eta \sqrt{\left(\frac{1}{v_\rho^2}+\frac{1}{v_\eta^2}\right) v_\chi^2+1}} \\
0 & v_\rho \sqrt{\frac{v_\eta^2+v_\chi^2}{\left(v_\rho^2+v_\chi^2\right) v_\eta^2+v_\rho^2 v_\chi^2}} & \frac{v_\eta v_\chi}{\sqrt{\left(v_\rho^2+v_\chi^2\right) v_\eta^2+v_\rho^2 v_\chi^2}} \\
\frac{v_\chi}{\sqrt{v_\eta^2+v_\chi^2}} & -\frac{v_\eta^2 v_\chi}{\sqrt{\left(v_\eta^2+v_\chi^2\right) \left(\left(v_\rho^2+v_\chi^2\right) v_\eta^2+v_\rho^2 v_\chi^2\right)}} & \frac{v_\eta v_\rho}{\sqrt{\left(v_\rho^2+v_\chi^2\right) v_\eta^2+v_\rho^2 v_\chi^2}} \\
\end{array}
\right).
\label{e1}
\end{equation}

The eigenvalues for the $2\times2$ part are $M^2_4$ and $M^2_5$:
\begin{equation}
\begin{array}{cl}
M^2_4&=-\frac{f_2 v_\rho v_\chi}{2 v_{s_2}}>0,\;\;f_2<0 \\
M^2_5&=-\frac{d_2 v_{s_2} v_\chi^2+2 f_2 v_\rho v_\chi}{4 v_{s_2}}>0,\;\; 2\vert f_2\vert v_\rho>d_2v_{s_2}.
\label{a2} 
\end{array}
\end{equation}
We see from Eqs.~(\ref{a1}) and (\ref{a2}), that the three physical pseudoscalar fields are heavy and also induce the type-II seesawlike mechanism in the charged lepton sector.

\subsection{Singly charged scalars 1}
\label{subsec:charged12}

If the lepton number is conserved in the scalar potential (\ref{potential}) under the conditions in Table~\ref{z7}, as we are assuming, the charged scalar mass matrix splits in two sectors $(\rho^+,\eta_1^+,h^+)$ and $(\chi^+, \eta_2^+, h_2^+)$. 

In the basis $(\rho^+,\eta_1^+,h^+)$, the mass matrix is
\begin{equation}
m^2_+\approx \frac{1}{2}\left(
\begin{array}{ccc}
-\frac{2 \sqrt{2} f_1 v_\eta v_\chi-2 a_9 v_\eta^2 v_\rho}{4 v_\rho} & \frac{f_1 v_\chi}{\sqrt{2}}-\frac{a_9 v_\eta v_\rho}{2} & 0 \\
\frac{f_1 v_\chi}{\sqrt{2}}-\frac{a_9 v_\eta v_\rho}{2} & \frac{1}{4} \left(2 a_9 v_\rho^2-\frac{2 \sqrt{2} f_1 v_\rho v_\chi}{v_\eta}\right) & 0 \\
0 & 0 & -\frac{f_2 v_\rho v_\chi}{2 v_{s_2}} \\
\end{array}
\right),
\end{equation}
with the following eigenvalues,
\begin{equation}
\begin{array}{cl}
m^2_{+1}&=0 \\
m^2_{+2}&= -\frac{f_2 v_\rho v_\chi}{2 v_{s_2}}>0,\;\;f_2<0 \\
m^2_{+3}&= \frac{\left(v_\eta^2+v_\rho^2\right) \left(a_9 v_\eta v_\rho-\sqrt{2} f_1 v_\chi\right)}{2 v_\eta v_\rho}\approx -\frac{1}{\sqrt2}\frac{f_1 v_\eta v_\chi}{v_\rho}>0,
\end{array}
\label{a3}
\end{equation}
and the eigenvectors are given by the column of the matrix
\begin{equation}
\left(
\begin{array}{ccc}
\frac{v_\rho}{\sqrt{v_\eta^2+v_\rho^2}} & 0 & -\frac{v_\eta}{\sqrt{v_\eta^2+v_\rho^2}} \\
\frac{v_\eta}{\sqrt{v_\eta^2+v_\rho^2}} & 0 & \frac{v_\rho}{\sqrt{v_\eta^2+v_\rho^2}} \\
0 & 1 & 0 \\
\end{array}
\right).
\end{equation}
According to (\ref{a3}), both physical charged scalar in this sector are very heavy. 

\subsection{Singly charged scalars 2}
\label{subsec:charged22}

In the other singly charged sector the mass matrix in the basis
$(\chi^+, \eta_2^+, h_2^+)$ is 
\begin{equation}
M_+^2 \approx \frac{1}{2} \left(
\begin{array}{ccc}
-\frac{2 \sqrt{2} f_1 v_\eta v_\rho-2 a_7 v_\eta^2 v_\chi}{4 v_\chi} & \frac{1}{2} \left(a_7 v_\eta v_\chi-\sqrt{2} f_1 v_\rho\right) & 0 \\
\frac{1}{2} \left(a_7 v_\eta v_\chi-\sqrt{2} f_1 v_\rho\right) & \frac{1}{2} v_\chi \left(a_7 v_\chi-\frac{\sqrt{2} f_1 v_\rho}{v_\eta}\right) & 0 \\
0 & 0 & -\frac{v_\chi (2 f_2 v_\rho+d_2 v_{s_2} v_\chi)}{4 v_{s_2}} \\
\end{array}
\right)
\end{equation}

the mass eigenvalues are,

\begin{equation}
\begin{array}{cl}
M^2_{+1}&=0 \\
M^2_{+2}&= -\frac{v_\chi (d_2 v_{s_2} v_\chi+2 f_2 v_\rho)}{4 v_{s_2}}>0, \;\;2\vert f_2\vert v_\rho>d_2v_{s_2}v_\chi\\
M^2_{+3}&= \frac{\left(v_\eta^2+v_\chi^2\right) \left(a_7 v_\eta v_\chi-\sqrt{2} f_1 v_\rho\right)}{2 v_\eta v_\chi}>0,
\label{a4}
\end{array}
\end{equation}
with the following eigenvectors:

\begin{equation}
\left(
\begin{array}{ccc}
-\frac{v_\chi}{\sqrt{v_\eta^2+v_\chi^2}} & 0 & \frac{v_\eta}{\sqrt{v_\eta^2+v_\chi^2}} \\
\frac{v_\eta}{\sqrt{v_\eta^2+v_\chi^2}} & 0 & \frac{v_\chi}{\sqrt{v_\eta^2+v_\chi^2}} \\
0 & 1 & 0 \\
\end{array}
\right).
\end{equation}
Again, the charged scalar masses in this sector might be heavy [see Eq.~(\ref{a4})].

\subsection{Double charge scalars}
\label{subsec:2charges}

The mass matrix in the basis $(\chi^{++},\rho^{++},S_1^{++},S_2^{++})$ is 
\begin{equation}
M_{++}^2 \approx \frac{1}{2}\left(
\begin{array}{cccc}
\frac{v_\rho \left(a_8 v_\rho v_\chi-\sqrt{2} f_1 v_\eta\right)}{2 v_\chi} & \frac{1}{2} \left(a_8 v_\rho v_\chi-\sqrt{2} f_1 v_\eta\right) & 0 & 0 \\
\frac{1}{2} \left(a_8 v_\rho v_\chi-\sqrt{2} f_1 v_\eta\right) & \frac{v_\chi \left(a_8 v_\rho v_\chi-\sqrt{2} f_1 v_\eta\right)}{2 v_\rho} & 0 & 0 \\
0 & 0 & -\frac{v_\chi (2 f_2 v_\rho+d_2 v_{s_2} v_\chi)}{4 v_{s_2}} & 0 \\
0 & 0 & 0 & \frac{v_\chi (d_2 v_{s_2} v_\chi-2 f_2 v_\rho)}{4 v_{s_2}} \\
\end{array}
\right),
\end{equation}
and the masses squared of the fields are,

\begin{equation}
\begin{array}{cl}
M^2_{++1}&=0 \\
M^2_{++2}&= \frac{v_\chi (d_2 v_{s_2} v_\chi-2 f_2 v_\rho)}{4 v_{s_2}}>0,\;\; \\
M^2_{++3} &= -\frac{v_\chi (d_2 v_{s_2} v_\chi+2 f_2 v_\rho)}{4 v_{s_2}}>0 \\
M^2_{++4}&= \frac{\left(v_\rho^2+v_\chi^2\right) \left(a_8 v_\rho v_\chi-\sqrt{2} f_1 v_\eta\right)}{2 v_\rho v_\chi}>0,
\end{array}
\end{equation}
with the respective eigenvectors:
\begin{equation}
\left(
\begin{array}{cccc}
-\frac{v_\chi}{\sqrt{v_\rho^2+v_\chi^2}} & 0 & 0 & \frac{v_\rho}{\sqrt{v_\rho^2+v_\chi^2}} \\
\frac{v_\rho}{\sqrt{v_\rho^2+v_\chi^2}} & 0 & 0 & \frac{v_\chi}{\sqrt{v_\rho^2+v_\chi^2}} \\
0 & 0 & 1 & 0 \\
0 & 1 & 0 & 0 \\
\end{array}
\right).
\end{equation}
In this doubly charged sector, all physical scalars might be heavy. 

\section{The lepton masses and the PMNS matrix}
\label{sec:pmns}

The problem of the leptonic mixing matrix has already been considered in the context of some 331 models without quarks with exotic charge, right-handed neutrinos  transforming nontrivially under $SU(3)_L$, three right-handed Majorana leptons, and with flavor  symmetries like $T_7$ \cite{Hernandez:2015cra}, $A_4$~\cite{Hernandez:2015tna} and extra scalars transforming as singlets under $SU(3)$. In these sort of models an $S_3$~\cite{Hernandez:2014lpa} symmetry is also used. Here, however, we will consider the m331 using only the gauge symmetries and also right-handed neutrinos as the extra degrees of freedom. 

As said before, the possibility that the lepton masses are generated by an effective dimensional operator was presented in Ref.~\cite{Montero:2001tq}. However, here we will consider the case when the sextet is introduced but its degrees of freedom are heavy enough to generate an effective nonrenormalizable interaction. This implements a type-II seesaw mechanism in the charged lepton sector. The latter situation arises when the members of the sextet are very heavy and one of the neutral scalars gains a zero VEV and the other one a small VEV. 

The Yukawa interactions are given by
\begin{eqnarray}
-\mathcal{L}^{lep}_1&=&-\frac{1}{2}\epsilon_{ijk}\,\overline{(\Psi_{ia})^c}G^\eta_{ab} \Psi_{jb}\eta_k+\frac{1}{2\Lambda_l}\,
\overline{(\Psi_{a})^c} \tilde{G}^s_{ab} (\chi^*\rho^\dagger+ \rho^*\chi^\dagger)\Psi_{b} \nonumber \\ &+&
\overline{(\Psi_{aL})}(G^\nu)_{ab}\nu_{aR}\eta+ \overline{(\nu_{aR})^c}(M_R)_{ab}\nu_{bR}+\overline{(\Psi_{ia})^c}G^s_{ab} \Psi_{jb}S_{ij}+H.c.,
\label{effective1}
\end{eqnarray}
where $\Lambda_l$ is a mass scale related to the origin of the dimension five interaction.
The second term in the first line of Eq.~(\ref{effective1}), the dimension-five operator, is generated by the loop in Fig.~\ref{fig1}. Notice that in Eq.~(\ref{effective1}) the interactions with the sextet appear and, although they do not contribute significantly to the charged lepton masses, the degrees of freedom in this multiplet might be exited at high energies.

We will assume, for the sake of simplicity, that $M_R$ is diagonal and that $m_{3R}\equiv M>m_{R1},m_{R2}$, and $M^{-1}_R=(1/M)\bar{M}_R$, where 
$\bar{M}=\textrm{diag}(r_{1},r_{2},1))$ and $r_1\equiv M/m_{R1},r_2\equiv M/m_{R2}$. In this case we have the mass matrices in the lepton sector
\begin{equation}
M^\nu\approx -\frac{v^2_\eta}{2}G^\nu\frac{\bar{M}_R}{M}G^{\nu T},\quad
M^l_{ab}=G^\eta_{ab}\frac{v_\eta}{\sqrt2}+\frac{1}{\Lambda_l}\tilde{G}^s_{ab}v_\rho v_\chi.
\label{mmassa3}
\end{equation}
If it is the sextet, through the interaction $\overline{(\Psi_{ai})^c} G^s_{ab} \Psi_{jb}S_{ij}$ and the $f_2$ trilinear term in the scalar potential involving $\rho,\chi$ and $S$, then  we have $1/\Lambda_l=f_2/m^2_{s_2} $ and
$\tilde{G}^s=G^s$. 

These mass matrices are diagonalized as follows: 
\begin{equation} 
\hat{M}^\nu=V^{\nu T}_LM^\nu V^\nu_L,\;\; \hat{M}^l=V^{l\dagger}_L M^l V^l_R,
\label{def}
\end{equation} 
where $\hat{M}^\nu = diag (m_1, m_2, m_3),\;\hat{M}^l = diag (m_e, m_\mu, m_\tau)$. The relation between symmetry eigenstates (primed) and mass eigenstates (unprimed) are $l^\prime_{L,R}=V^l_{L,R}l_{L,R}$
and $\nu^\prime_L=V^\nu_L \nu_L$, where
$l^\prime_{L,R}=(e^\prime,\mu^\prime,\tau^\prime)^T_{L,R}$, $l_{L,R}=(e,\mu,\tau)^T_{L,R}$  and
$\nu ^\prime_L=(\nu_e\,\nu_\mu\,\nu_\tau)^T_L$
and $\nu_L=(\nu_1\,\nu_2\,\nu_3)_L$. 

In the following we assume  $v_\chi\approx \Lambda_l$ and, as in Ref.~\cite{Machado:2013jca}, $v_\rho\sim 54,v_\eta\sim 240$ GeV. 
The neutrino mass matrix is as in Eq.~(\ref{mmassa3}). Solving simultaneously the following equations:
\begin{equation}
\hat{M}^\nu_{ab}=V^{\nu T}_L M^\nu_{ab}V^\nu_L,\quad 
V^{l\dagger}_L M^l M^{l\dagger}V^l_L=V^{l\dagger}_R M^{l\dagger}  M^lV^l_R=(\hat{M}^l)^2,\quad V_{PMNS}=V^{l\dagger}_LV^\nu_L,
\end{equation}
where $M^\nu$ and $M^l$ are defined in Eq.~(\ref{mmassa3}), and $V_{PMNS}$ is the mixing matrix in the lepton sector (PMNS), the values for the charged lepton masses obtained are (in MeV) $(m_e,m_\mu,m_\tau)=(0.509648,105.541,1775.87)$
and for the neutrinos masses (in eV) $(m_1,m_2,m_3)=(0.051,-0.0194,0.0174)$ which are consistent with 
$\Delta m^2_{23}=2.219\times10^{-3}\,(\textrm{eV})^2$ and $\Delta m^2_{21}=7.5\times 10^{-5}\,(\textrm{eV})^2$. 
These values for the masses arise from the following values for the Yukawa matrices: $v^2_\eta/M=0.33$ eV, 
$G^\nu_{11}=0.109$, $G^\nu_{12}=0.097$, 
$G^\nu_{13}=0.101$, $G^\nu_{22}=0.09$, $G^\nu_{23}=-0.02$, $G^\nu_{33}=0.0106$ in the neutrino sector; and 
$G^s_{11}=-0.0453,G^s_{12}=-0.0076,G^s_{13}=-0.0008,G^s_{22}=0.0015,
G^s_{23}=0.0001,G^s_{33}=1.84\times10^{-5}$,
$G^\eta_{12}=,G^\eta_{13}=G^\eta_{13}=-0.00001$ in the charged lepton sector. The only way to avoid the latter fine-tunning is 
to consider $v_\eta$ smaller, but in the context of the Ref.~\cite{Machado:2013jca} this VEV is already fixed.

We obtain for the diagonalization matrices
\begin{equation}
V^\nu_L\approx\left(\begin{array}{ccc}
-0.24825& -0.57732& 0.77786\\
0.73980 &-0.40539  &0.53698 \\
-0.62535 &-0.70877 &0.32647 \\
\end{array}\right)
\label{lep1}
\end{equation}
and
\begin{equation}
V^l_L\approx\left(\begin{array}{ccc}
-0.00985 &0.01457 &-0.99984 \\
-0.31848& -0.94787 &-0.01067 \\
0.94788&-0.31833 & -0.01398\\
\end{array}\right),\quad 
V^l_R\approx\left(\begin{array}{ccc}
0.00501&0.00716 & 0.99996\\
0.00261&0.9910  & -0.00717\\
0.99998 &-0.00265 & -0.00499\\
\end{array}\right)
\label{lep2}
\end{equation}

Notice that we have defined the lepton mixing matrix as $V_{PMNS}=V^{l\dagger}_LV^\nu_L$, which means that this matrix appears in the charged currents coupled to $W^-$.  We obtain from Eqs.~(\ref{lep1}) and (\ref{lep2}) the following values for the PMNS matrix:
\begin{equation}
\vert V_{PMNS}\vert \approx\left(\begin{array}{ccc}
0.826& 0.548 & 0.130\\
0.506&0.618  &0.602 \\
0.249 &0.563 &0.788 \\
\end{array}\right),
\label{pmns}
\end{equation}
which is in agreement, within 3$\sigma$, with the experimental data given in Ref.~\cite{GonzalezGarcia:2012sz},
\begin{equation}
\vert V_{PMNS}\vert \approx\left(\begin{array}{ccc}
0.795-0.846& 0.513-0.585 & 0.126-0.178\\
0.205-0.543&0.416-0.730  &0.579 - 0.808 \\
0.215 - 0.548 &0.409 - 0.725 &0.567 -0.800 \\
\end{array}\right),
\label{pmnsexp}
\end{equation}
and we see that it is possible to accommodate all lepton masses and the PMNS matrix. We do not consider $CP$ violation here.

\section{Conclusions}
\label{sec:con}

We have shown that even if we introduce the sextet in such a way that it practically does not contribute to the lepton masses because of the small VEVs, and since its components are very heavy, it might generate the dimension-five operator involving only the triplets $\rho$ and $\chi$ in Eq.~(\ref{effective1}) trough a process like the one shown in Fig.~\ref{fig1}. A similar operator can be obtained for the neutrinos as in Ref.~\cite{Montero:2001tq}, but here we prefer to introduce right-handed neutrinos in order to implement a type-I seesaw mechanism. Moreover, in the charged lepton sector the mass generation is similar to the type-II seesaw mechanism inducing small masses for neutrinos through the exchange of a heavy non-Hermitian triplet~\cite{Cheng:1980qt,Konetschny:1977bn,Magg:1980ut}.
The existence of several mechanisms to generate this interaction in the context of the standard model have been shown in the literature~\cite{Ma:1998dn,Bonnet:2012kz,Sierra:2014rxa}.  Notwithstanding, the effective operator in Eq.~(\ref{effective1}) can be originated by the effects of higher-dimension operators. 

In our case, the sextet is introduced in the model as in the m331 model, and through its interactions with the other scalars in the scalar potential and with leptons in the Yukawa interactions, their degrees of freedom can be exited at high energies mainly in lepton colliders.
Notice, that all extra scalars in the model are heavy except for two neutral scalars that correspond to the fields in a two-Higgs-doublet extension of the SM.
We also show the conditions under which we have a weak copositivity of the scalar potential and generate the vacuum stability at tree level and a global minimum as well. It is interesting to study the same problem at the one-loop level. For instance, in a model with two doublets (one of them inert) taking into account a neutral scalar with a mass of 125 GeV, the stability of the vacuum was shown in Ref.~\cite{Goudelis:2013uca}; however, such analysis is beyond the scope of our paper.

In order to obtain the correct mass for the charged leptons, besides the effective interaction, it is necessary to consider the interactions with the triplet $\eta$. For neutrinos, as we are considering that $v_{s_1}=0$, we have introduced right-handed components to generate the type-I seesaw mechanism. With the unitary (orthogonal if we neglect phases) matrices that diagonalize the mass matrices in the lepton sector, it is possible to accommodate a realistic Pontecorvo-Maki-Nakawaga-Sakata matrix. The constraints on the masses of the extra particles in the m331 model coming from lepton violation processes will be considered elsewhere.
In the present context we recall that it is the neutral scalar $\rho^0$ which has the larger projection on the neutral scalar with a mass of near 125 GeV. For details see Ref.~\cite{Machado:2013jca}.

Finally, we would like to discuss the differences between our present model and those in Refs.~\cite{Montero:2001tq,Ferreira:2011hm}. Our model is the usual minimal 3-3-1 (in the sense that the lepton sector consists only of the known leptons) and the four scalar multiplets, three triplets, and one sextet \textit{plus} right-handed neutrinos. Although the sextet has interactions mediated by scalars [see Eq.~(\ref{effective1})], its neutral components do not contribute significantly to the lepton masses.
The degrees of freedom in the sextet decouple at low energies; however, its interactions have to be taken into account for some phenomenology at sufficiently high energy. For instance, if this mechanism is implemented at the 100 GeV--1TeV scale, the interactions of the charged leptons with the left-handed neutrinos could have some signature at the LHC or other colliders. (See \cite{Ng:2015hba} and references therein.)

In  Ref.~\cite{Montero:2001tq}, only three triplets $\eta,\rho$ and $\chi$ were considered, being the sextet avoided at all, with the charged leptons and neutrinos gain mass only through nonrenormalizable interactions. The model of Ref.~\cite{Ferreira:2011hm} is  considered the so called "reduced" 3-3-1 model with only the triplets $\rho$ and $\chi$ (with no $\eta$ and $S$ at all). Therefore, for generating all the fermion masses they need, besides the usual Yukawa interactions with these triplets, nonrenormalizable interactions involving the same triplets. Although this is an interesting situation, the model with only the triplets $\rho$ and $\chi$ has experimental troubles if we accept the existence of the Landau-like pole. See Ref.~\cite{Dong:2014bha} for a discussion of the 3-3-1 models with only two triplets, in particular the troubles with the ``reduced" minimal 3-3-1 model~\cite{Ferreira:2011hm}.   

Thus, in the limit of a heavy sextet, our model is a 3-3-1 model with three triplets and the sextet, but the latter one does not contribute to the lepton masses and almost not at all to the spontaneous symmetry breaking since its VEVs are zero or very small in the sense of $v_{s_2}/v_W\ll1$. The lepton interactions with the sextet and the interactions among all the scalars 
become important only at high energies. Our case is more similar to the type-II seesaw mechanism in which a complex heavy triplet is in charge of generating the neutrino masses~\cite{Ma:1998dn}.

\acknowledgments

The authors would like to thank for full support to CNPq  (GDC), CAPES (ACBM) and for partial
support to CNPq (VP).

\appendix

\section{Stability of the scalar potential}
\label{sec:appendixa}

The scalar potential has to be bounded from below to ensure its stability. In the SM this is easy: we
just have to ensure that $\lambda > 0$ (this also ensure the existence of a global minimum). In theories that increase the number of scalars it is more difficult to ensure that the potential is
bounded from below in all directions. The following cases are based on the copositivity matrix presented in Eq.~(\ref{matriz44}) below. Each case corresponds to different sign possibilities of the matrix elements. Also, for every case, all diagonal elements have to be positive. Here we follow Refs.~\cite{ping,Kannike:2012pe}. The copositivity criteria only assure that the potential is bounded from below. In order to verify if there exist a  global minimum, it must be be calculated only numerically. This is very complicated even for the case of the SM extensions with two  \cite{Maniatis:2006fs} and three \cite{Maniatis:2014oza} scalar doublets. However, we show below that there is a global minimum in the scalar potential if some reasonable conditions are imposed.

A scalar potential has a quadratic form in the quadratic couplings in the form $A_{ab}
\phi^2_a \phi^2_b $, if the matrix $A_{ab}$ is copositive it is possible to ensure that the potential has a global minimum. To do this analysis, 
we can ignore any terms with couplings with dimension, mass or soft terms, since in the limit of large field values terms with dimension smaller than four are negligible in comparison with the quartic part of the scalar potential $V^{4}$. 
In addition, we will assume the conditions from Table \ref{z7}, so that $f_3,f_4,a_{10},b_1,b_2,b_3,c_1,c_2=0$. Taking the potential in Eq.~(\ref{potential2}) we redefine the triplets and the sextet as:
\begin{equation}
|H_i|^2=h_i^2, \qquad H_i^\dagger H_j = h_i h_j r_{ij} e^{i \theta_{ij}}
\label{eq:copositivity}
\end{equation}
where $H_{1,2,3,4}=\eta,\rho,\chi,S$. The parameters $r_{ij}$ and $\theta_{ij}$ are not physical parameters, ranging from 0 to 1 and $0$ to $2\pi$, respectively. They  are used to analyze the four-field direction by demanding the maximization of the parameter space. Then, they should be set to values which allow the most parameter space.

Rewriting the quartic terms of the potential in Eq.~(\ref{potential2}), and  using Eqs. (\ref{eq:copositivity}) and Table~\ref{z7},  we find that the matrix $A$ in the basis  $(h_1^2,h_2^2,h_3^2,h_4^2)$ is written as

\begin{equation}
A=\left(
\begin{array}{cccc}
a_1 & a_6+a_9r^2_{12} & a_4+a_7r^2_{13} & d_3+d_4 r^2_{14}\\
a_6+a_9r^2_{12} & a_2 & a_5+a_8r^2_{23} & d_5+d_6 r^2_{24}\\
a_4+a_7 r^2_{13} & a_5+a_8r^2_{23} & a_3 & d_1+d_2r^2_{34}\\
d_3+ d_4 r^2_{14} & d_5+d_6r^2_{24} & d_1+d_2r^2_{34}& e_1+e_2
\end{array}
\right).
\label{matriz44}
\end{equation}

In order to analyze the copositivity of the matrix $A$ above we have to choose values for the $r_{ij}$ that minimize the entries of the matrix. For the off-diagonal elements, which involve sums, two cases are relevant: if both coupling constants are positive/negative, the minimum comes from choosing $r_{ij}=0$; if the constants have opposite signs, the minimum comes from $r_{ij}=1$. For the sake of simplicity, we will assume all coupling constants on each entry to have the same sign; therefore, we make all $r_{ij}=0$ and consider six cases for the copositivity.

Below are shown the different conditions under which the potential is bounded from below.
All indices $i,j,k,l,...$ are fixed and different from each other. Also, we want to remind the reader that all diagonal elements from matrix A given in Eq.~(\ref{matriz44}) should be positive to have copositivity.

\noindent  \textbf{Case 1.} All $A_{ij}$ positive.\\
All couplings positive, and $E=e_1+e_2 >0$.

\noindent  \textbf{Case 2.} $A_{ij}\leq 0$ with $i,j$ and the other entries positive.
\begin{enumerate}
	\item If $a_6\leq0$ we have $a_1a_2-a^2_6>0$.
	\item If $a_4\leq 0$, then $a_1a_3-a^2_4>0$.
	\item If $a_5\leq0$, then $a_2a_3-a^2_5>0$.
	\item If $d_3\leq 0$, then $a_1 E - d_3^2>0$.
	\item If $d_5\leq0$, then $-d^2_5+a_2E>0$ .
	\item If $d_1\leq0$, then $-d^2_1+a_3E>0$.
\end{enumerate}

\noindent  \textbf{Case 3.} $A_{ij}\leq0$, $A_{kl}\leq 0$, and the other entries positive. 
\begin{enumerate}
	\item If $a_5\leq0$ and $a_6\leq0$, we have $a_1a_2-a^2_6>0$ and $a_2a_3-a^2_5>0$.
	\item If $a_6\leq 0$ and $d_5\leq0$, then $a_1a_2-a^2_6>0$ and $-d^2_5+a_2E>0$.
	\item If $a_6\leq0$ and $d_1\leq0$, then $a_1a_2-a^2_6>0$ and $-d^2_1+a_3E>0$.
	\item If $a_4\leq0$ and $d_5\leq0$, then $a_1a_3-a^2_4>0$ and $-d^2_5+a_2E>0$.
	\item If $a_4\leq0$ and $d_1\leq0$, then $a_1a_3-a^2_4>0$ and $-d^2_1+a_3E>0$.
	\item If $a_5\leq0$ and $d_3\leq0$, then $a_2a_3-a^2_5>0$ and $-d_3^2+a_1E>0$.
	\item If $a_5\leq0$ and $d_1\leq0$, then $a_2a_3-a^2_5>0$ and $-d_1^2+a_3E>0$.
\end{enumerate}

\noindent \textbf{Case 4} As in case 3: $A_{ij}\leq0$, $A_{ik}\leq0$ 
and the other entries positive
\begin{enumerate}
	\item If $a_4\leq0$ and $a_6\leq0$, then $a_1a_5+\sqrt{(a_1a_3-a^2_4)(a_1a_2-a^2_6)}>0$.
	\item If $a_6\leq0$ and $d_3\leq0$, then $a_1d_5+\sqrt{(a_1a_2-a^2_6)(-d_3^2+a_1E)}>0$.
	\item If $a_4\leq0$ and $d_3\leq0$, then $a_1d_1+\sqrt{(a_1a_3-a^2_4)(-d_3^2+a_1E)}>0$.
	\item If $a_5\leq0$ and $d_5\leq0$, and $a_2d_1+\sqrt{(a_2a_3-a^2_5)[-d^2_5+a_2E]}>0$.
\end{enumerate}

\noindent \textbf{Case 5.} $A_{ij}\leq0$, $A_{jk}\leq0$, $A_{ik}\leq0$ and the other entries positive
\begin{enumerate}
	\item If $a_4,a_5,a_6\leq0$, then 
	\begin{eqnarray}
	&&\sqrt{a_1a_2}+a_6>0,\quad \sqrt{a_1a_3}+a_4>0,\quad \sqrt{a_2a_3}+a_5>0,\nonumber \\&&
	a_1a_2a_3-a_2a^2_4-a_1a^2_5+2a_4a_5a_6-a_3a^2_6>0
	\label{c51}
	\end{eqnarray} 
	\item If $a_6,d_5,d_3\leq0$, then 
	\begin{eqnarray}
	&& \sqrt{a_1a_2}+a_6>0,\quad d_3+\sqrt{a_1E}>0,\quad 
	d_5+\sqrt{a_2E}>0,\nonumber \\&&
	-a_2 d_3^2+2a_6d_3d_5-a_1d_5^2+(a_1a_2-a_6^2)E>0
	\label{c52}
	\end{eqnarray}
	
	\item If $a_4,d_1,d_3\leq0$,then
	\begin{eqnarray}
	&& \sqrt{a_1a_3}+a_4>0,\quad d_3+ \sqrt{a_1E}>0,\quad 
	d_1+\sqrt{a_3E}>0,\nonumber \\&&
	2a_4d_1d_3-a_3d_3^2+(a_1 a_3 - a_4^2)E-a_1 d_1^2>0
	\label{c53}
	\end{eqnarray}
	
	\item If $a_5,d_1,d_5\leq0$,then
	\begin{eqnarray}
	&& \sqrt{a_2a_3}+a_5>0,\quad d_5+\sqrt{a_2E}>0,\quad 
	d_1+\sqrt{a_3E}>0,\nonumber \\&&
	-a_2 d_1^2+2 a_5 d_1 d_5-a_3 d_5^2+a_2 a_3 e_1-a_5^2 e_1+a_2 a_3 e_2-a_5^2 e_2>0,\nonumber \\&& 
	\label{c54}
	\end{eqnarray}
	
\end{enumerate}

\noindent \textbf{Case 6.} $A_{ij}\leq0$, $A_{ik}\leq0$, $A_{il}\leq0$, and the other entries positive

\begin{enumerate}
	\item If $a_4,a_6,d_3\leq0$, then
	\begin{eqnarray}
	&&a_1a_2-a^2_6>0,\quad a_1a_3-a^2_4>0,\quad-d_3^2+a_1E>0,\nonumber \\&&
	a_1a_5-a_4a_6+\sqrt{(a_1a_3-a^2_4)(a_1a_2-a^2_6)}>0,
	\nonumber \\&&
	a_1d_5+\sqrt{(a_1a_2-a^2_6)[-d_3^2+a_1E]}>0,\nonumber \\&&
	a_1d_1+\sqrt{(a_1a_3-a^2_4)[-d_3^2+a_1E]}>0,\nonumber\\&&
	-a_1^2 \lbrace -2a_2a_4d_1d_3+a_2a_3d_3^2-a_5^2d_3^2+2a_4a_5d_3d_5-a_4^2d_5^2 \nonumber\\&& +a_2a_4^2E+a_6^2(-d_1^2+a_3E)+2a_6(a_5d_1d_3+a_4d_1d_5-a_3d_3d_5-a_4a_5E)\nonumber\\&&+a_1(-2a_5d_1d_5+a_3d_5^2+a_5^2E+a_2(d_1^2-a_3E)) \rbrace >0
	\label{c61}
	\end{eqnarray}
	
	\item If $a_5,a_6,d_5\leq0$, then
	\begin{eqnarray}
	&&a_1a_2-a^2_6>0,\quad a_2a_3-a^2_5>0,\quad-d_5+a_2E>0,\nonumber \\&&
	a_2a_4-a_5a_6+\sqrt{(a_2a_3-a^2_5)(a_1a_2-a^2_6)}>0,
	\nonumber \\&-&
	a_2^2 \lbrace-a_6^2 d_1^2 + 2 a_4 a_6 d_1 d_5 + a_1 a_3 d_5^2 - 
	a_4^2 d_5^2 + a_3 a_6^2 E
	\nonumber \\&+&
	a_5^2  a1 E + 
	2 a_5 ( - a_1 d_1 d_5 - a_4 a_6 E)
	\nonumber \\&+& 
	a_2 [ + a_1 d_1^2 + a_4^2 E - a_3  a_1 E]\rbrace>0
	\label{c62}
	\end{eqnarray}
	
	\item If $a_4,a_5,d_1\leq0$, then
	\begin{eqnarray}
	&&a_1a_3-a^2_4>0,\quad a_2a_3-a^2_5>0,\quad-d_1+a_3E>0,\nonumber \\&&
	-a_4a_5+a_3a_6+\sqrt{(a_1a_3-a^2_4)(a_2a_3-a^2_5)}>0,
	\nonumber \\&&
	-a_4d_1+\sqrt{(a_1a_3-a^2_4)[-d^2_1+a_3 E]}>0,
	\nonumber \\&-&
	a_3^2 \lbrace-a_6^2 d_1^2 + 2 a_4 a_6 d_1 d_5 + a_1 a_3 d_5^2 - 
	a_4^2 d_5^2 + a_3 a_6^2 E
	\nonumber \\&+&
	a_5^2  a_1 E + 
	2 a_5 ( - a_1 d_1 d_5 - a_4 a_6 E)
	\nonumber \\&+&
	a_2  + a_1 d_1^2 + a_4^2 E- a_3 a_1 E)]\rbrace>0
	\label{c63}
	\end{eqnarray}
	
	\item If $d_1,d_3,d_5\leq0$, then
	\begin{eqnarray}
	&&-d_3^2+a_1E>0,\quad -d^2_5+a_2E>0,\quad-d^2_1+a_3E>0,\nonumber \\&&
	a_6E+\sqrt{[-d_3^2+a_1E][-d^2_5+a_2E]}>0,
	\nonumber \\&&
	a_4E+\sqrt{[-d_3^2+a_1E][-d^2_1+a_3E]}>0,\nonumber \\&&
	-d_1d_5+a_5E+\sqrt{[-d^2_5+a_2E][-d^2_1+a_3E]}>0,
	\nonumber \\&&
	E^2 \lbrace 2a_2a_4d_1d_3-a_2a_3d_3^2+a_5d_3^2-2a_4a_5d_3d_5+a_4^2d_5^2-a_2a_4^2E \nonumber \\&& +a_6^2(d_1^2-a_3E)+2a_6(-a_5d_1d_3-a_4d_1d_5+a_3d_3d_5+a_4a_5E) \nonumber \\&& -a_1(-2a_5d_1d_5+a_3d_5^2+a_5^2E+a_2(d_1^2-a_3E)) \rbrace>0 \nonumber\\&&
	\label{c64}
	\end{eqnarray}
	
\end{enumerate}

The potential at the minimum is
\begin{eqnarray}
V & = & \frac{1}{2} \left(\mu_{\eta}^2 v_{\eta}^2 + \mu_{\rho}^2 v_{\rho}^2 + \mu_{\chi}^2 v_{\chi}^2 + \mu_{S}^2 v_{s_2}^2 \right)  + \frac{1}{2 \sqrt{2}} \left(f_1 v_\eta v_\rho v_\chi + f_2 v_2 v_\rho v_\chi   \right) \nonumber \\ &+& \frac{1}{4} \left( 2 v_{s_2}^4 ( 2e_1 + e_2) + v_{\eta}^2 (2 d_3 v_{s_2}^2 + a_6 v_{\rho}^2 + a_4 v_{\chi}^2 + a_1 v_{\eta}^2) \right)   \nonumber \\ &+& \frac{1}{4} \left( v_{\chi}^2 ((2 d_1 + d_2) v_{s_2}^2 + a_5 v_\rho^2 + a_3 v_{\chi}^2)  + v_{\rho}^2 ((2 d_5 + d_6) v_{s_2}^2) \right),
\label{minimum}
\end{eqnarray}
and since all VEVs are positive and $\mu_X^2,f_1,f_2<0$, if we impose that $a_4<0$, $a_3\ll \vert a_4\vert$ and $a_6,a_1,a_5<\vert a_4\vert$, the potential is at a global minimum independent of the sign of the other terms in  (\ref{minimum}), once they are negligible. 

\section{Goldstone bosons}
\label{sec:appendixb}

In this appendix we present all the mass matrices for the scalars in the potential without any approximation. When analytical results are available for the mass eigenvalues they are presented right below the mass matrix.  We also present the Goldstone eigenvectors for the applicable matrices.

\subsection{Neutral CP-odd scalars}
\label{subsec:cpodd}

Mass matrix $M^2_I$, in the basis $(I_\eta^0\,I_\rho^0\, I_\chi^0\, I_{s2}^0 )$, is given by
\begin{equation}
\left(
\begin{array}{cccc}
-\frac{f_1 v_\rho v_\chi}{\sqrt{2} v_\eta} & -\frac{f_1 v_\chi}{\sqrt{2}} & -\frac{f_1 v_\rho}{\sqrt{2}} & 0 \\
& -\frac{\left(\sqrt{2} f_1 v_\eta+f_2 v_{s_2}\right) v_\chi}{2 v_\rho} & -\frac{f_1 v_\eta}{\sqrt{2}}-\frac{f_2 v_{s_2}}{2} & -\frac{1}{2} f_2 v_\chi \\
 &  & -\frac{v_\rho \left(\sqrt{2} f_1 v_\eta+f_2 v_{s_2}\right)}{2 v_\chi} & -\frac{1}{2} f_2 v_\rho \\
 &  &  & -\frac{f_2 v_\rho v_\chi}{2 v_{s_2}} 
\end{array}
\right).
\label{i2}
\end{equation}

The mass eigenvalues are denoted by $m^2_i,\;i=1,\cdots,5$.
This matrix also decomposes into $4\times4$ + $1\times1$, and it has the following eigenvalues. 
The eigenvalue for the 1$\times$1 part is denoted by $\lambda_5$ and is presented below.

\begin{equation}
\begin{array}{cl}
M^2_1= & M^2_2=0\\
M^2_3=&-\frac{1}{{4 v_\eta v_\rho v_{s_2} v_\chi}}\Big\lbrace \Big[ \left[\sqrt{2} f_1 v_{s_2} [v_\eta^2 (v_\rho^2+v_\chi^2)+v_\rho^2 v_\chi^2]+f_2 v_\eta [v_\rho^2 \left(v_{s_2}^2+v_\chi^2\right)+v_{s_2}^2 v_\chi^2]\right]^2 \\ & 
-4 \sqrt{2} f_1 f_2 v_\eta v_\rho^2 v_{s_2} v_\chi^2 [v_\chi^2 v^2_W +v_\rho^2 (v_\eta^2+v_{s_2}^2)] \Big]^{\frac{1}{2}}  \\ &  +\sqrt{2} f_1 v_{s_2} \left(v_\eta^2 \left(v_\rho^2+v_\chi^2\right)+v_\rho^2 v_\chi^2\right)+f_2 v_\eta \left(v_\rho^2 \left(v_{s_2}^2+v_\chi^2\right)+v_{s_2}^2 v_\chi^2\right)\Big\rbrace \\
M^2_4=&\frac{1}{4 v_\eta v_\rho v_{s_2} v_\chi} \Big\lbrace \Big[ [\sqrt{2} f_1 v_{s_2} \left(v_\eta^2 \left(v_\rho^2+v_\chi^2\right)+v_\rho^2 v_\chi^2]+f_2 v_\eta \left(v_\rho^2 \left(v_{s_2}^2+v_\chi^2\right)+v_{s_2}^2 v_\chi^2\right)\right)^2 \\ & -4 \sqrt{2} f_1 f_2 v_\eta v_\rho^2 v_{s_2} v_\chi^2 [v_\chi^2 v^2_W+v_\rho^2 \left(v_\eta^2+v_{s_2}^2\right)] \Big]^{\frac{1}{2}} \\ & -\sqrt{2} f_1 v_{s_2}[v_\eta^2 \left(v_\rho^2+v_\chi^2\right)+v_\rho^2 v_\chi^2]-f_2 v_\eta [v_\rho^2 \left(v_{s_2}^2+v_\chi^2\right)+v_{s_2}^2 v_\chi^2] \Big\rbrace \\
M^2_5=&\frac{1}{4 v_{s_2}}(d_2 v_{s_2} v_\chi^2+2 d_4 v_\eta^2 v_{s_2}-d_6 v_\rho^2 v_{s_2}-2 e_2 v_{s_2}^3-2 f_2 v_\rho v_\chi).
\end{array}
\end{equation}
where we have defined $v^2_W=v_\eta^2+v_\rho^2+v_{s_2}^2$. We write explicitly only the 
Goldstone eigenvectors:
\begin{equation}
G_1^0=\frac{1}{\sqrt{v_\eta^2+v_\chi^2+v_{s_2}^2}}\left(v_\eta,\; 0, \;-v_\chi,\; v_{s_2}, \, 0\right)^T
\end{equation}

\begin{equation}
G_2^0=\frac{1}{\sqrt{v_\chi^2 v^2_W +v_\rho^2 \left(v_\eta^2+v_{s_2}^2\right)}}
\left(
\begin{array}{c}
-\frac{v_\eta v_\chi^2}{\sqrt{\left(v_\eta^2+v_{s_2}^2+v_\chi^2\right)}} \\
v_\rho \sqrt{v_\eta^2+v_{s_2}^2+v_\chi^2}\\
-\frac{v_\chi \left(v_\eta^2+v_{s_2}^2\right)}{\sqrt{\left(v_\eta^2+v_{s_2}^2+v_\chi^2\right)}}\\
-\frac{v_{s_2} v_\chi^2}{\sqrt{\left(v_\eta^2+v_{s_2}^2+v_\chi^2\right)}}\\
0
\end{array}
\right)
\end{equation}

\subsection{Singly charged scalars 1}
\label{subsec:charged1}

The mass matrix $M^2_{c1}$, with the basis $(\rho^+\,\eta_1^+\,h^+)$ given by
\begin{equation}
\left(
\begin{array}{ccc}
\frac{(2 a_9  v_\eta^2-d_6  v_{s_2}^2)v_\rho-2 \left(\sqrt{2} f_1 v_\eta+f_2 v_{s_2}\right) v_\chi}{4 v_\rho} & \frac{f_1 v_\chi}{\sqrt{2}}-\frac{a_9 v_\eta v_\rho}{2} & \frac{1}{4} (d_6 v_\rho v_{s_2}+2 f_2 v_\chi) \\
 & \frac{1}{4} \left(2 a_9 v_\rho^2-\frac{2 \sqrt{2} f_1 v_\chi v_\rho}{v_\eta}+d_4 v_{s_2}^2\right) & -\frac{1}{4} d_4 v_\eta v_{s_2} \\
 &  & \frac{d_4 v_\eta^2 v_{s_2}-v_\rho (d_6 v_\rho v_{s_2}+2 f_2 v_\chi)}{4 v_{s_2}} \\
\end{array}
\right)
\end{equation}
\\
The mass eigenvalues
\begin{equation}
m^2_{+1}= 0,\quad m^2_{+2}=-\frac{1}{8 v_\eta v_\rho v_{s_2}}(A+B),\quad 
m^2_{+3}=-\frac{1}{8 v_\eta v_\rho v_{s_2}}(A-B),
\end{equation}
where 
\begin{eqnarray}
&& A=v_\eta v_\rho v_{s_2}	[ (2 a_9 (v_\eta^2+v_\rho^2)+d_4(v_\eta^2+v_{s_2}^2)-d_6 (v_\rho^2+v_{s_2}^2)) \nonumber \\&& 
-2 v_\chi (\sqrt{2} f_1 v_{s_2}(v_\eta^2+v_\rho^2)+f_2 v_\eta (v_\rho^2+v_{s_2}^2)))^2 \nonumber \\ && -4v^2_W (2 a_9 v_\eta v_\rho (d_4 v_\eta^2 v_{s_2}-v_\rho (d_6 v_\rho v_{s_2}+2 f_2 v_\chi)) \nonumber \\ &&  +2 \sqrt{2} f_1 v_\chi (v_\rho (d_6 v_\rho v_{s_2}+2 f_2 v_\chi)-d_4 v_\eta^2 v_{s_2})-d_4 v_\eta v_{s_2}^2 (d_6 v_\rho v_{s_2}+2 f_2 v_\chi)) \Big]^{\frac{1}{2}}\nonumber \\ &&
B=v_\eta v_\rho v_{s_2}[-2 a_9  (v_\eta^2+v_\rho^2)-d_4 (v^2_\eta- v_{s_2}^2)+
d_6 (  v_\rho^2 + v_{s_2}^2) ]\nonumber \\&& +2v_\chi[ \sqrt{2}f_1  v_{s_2}(v_\eta^2 + v_\rho^2) + f_2 v_\eta (v_\rho^2 +  v_{s_2}^2)]
\end{eqnarray} 	
The Goldstone eigenvector is

\begin{equation}
G_1^-=\frac{1}{\sqrt{v_\rho^2+v_\eta^2+v_{s_2}^2}}(v_\rho, \, v_\eta, \, v_{s_2})^T
\end{equation}

\subsection{Singly charged scalars 2}
\label{subsec:charged2}

The mass matrix  is $M^2_{c2}$, with the basis $(\chi^+\, \eta_2^+\, h_2^+)$ given by
\begin{equation}
\left(
\begin{array}{ccc}
\frac{2 a_7 v_\chi v_\eta^2-2 \sqrt{2} f_1 v_\rho v_\eta-2 f_2 v_\rho v_{s_2}-d_2 v_{s_2}^2 v_\chi}{4 v_\chi} & \frac{1}{2} \left(a_7 v_\eta v_\chi-\sqrt{2} f_1 v_\rho\right) & \frac{1}{4} (2 f_2 v_\rho+d_2 v_{s_2} v_\chi) \\
 & \frac{1}{4} \left(d_4 v_{s_2}^2+2 v_\chi \left(a_7 v_\chi-\frac{\sqrt{2} f_1 v_\rho}{v_\eta}\right)\right) & \frac{1}{4} d_4 v_\eta v_{s_2} \\
 &  & \frac{d_4 v_\eta^2 v_{s_2}-v_\chi (2 f_2 v_\rho+d_2 v_{s_2} v_\chi)}{4 v_{s_2}} \\
\end{array}
\right)
\end{equation}
\\
The mass eigenvalues are
\begin{equation}
M^2_{+1}=0,\quad M^2_{+2}=-\frac{1}{8 v_\eta v_{s_2} v_\chi} [A^\prime +B^\prime],\quad 
M^2_{+2}=-\frac{1}{8 v_\eta v_{s_2} v_\chi} [A^\prime -B^\prime]
\end{equation}
where

\begin{eqnarray}
&& A^\prime= [(v_\eta (2 f_2 v_\rho (v_{s_2}^2+v_\chi^2)-v_{s_2} v_\chi (2 a_7 (v_\eta^2+v_\chi^2)-d_2 (v_{s_2}^2+v_\chi^2)+d_4 (v_\eta^2+v_{s_2}^2)))\nonumber \\ && +2 \sqrt{2} f_1 v_\rho v_{s_2} (v_\eta^2+v_\chi^2))^2+4 v_\eta v_{s_2} v_\chi (v_\eta^2+v_{s_2}^2+v_\chi^2) (d_4 v_\eta v_{s_2} (-2 a_7 v_\eta^2 v_\chi+d_2 v_{s_2}^2 v_\chi \nonumber \\ && +2 \sqrt{2} f_1 v_\eta v_\rho+2 f_2 v_\rho v_{s_2})+2 v_\chi (a_7 v_\eta v_\chi-\sqrt{2} f_1 v_\rho) (d_2 v_{s_2} v_\chi+2 f_2 v_\rho)) ]^{\frac{1}{2}} \nonumber \\ && B^\prime= -v_\eta (v_{s_2} v_\chi (2 a_7 (v_\eta^2+v_\chi^2)-d_2 (v_{s_2}^2+v_\chi^2)+d_4 (v_\eta^2+v_{s_2}^2))-2 f_2 v_\rho (v_{s_2}^2+v_\chi^2))\nonumber \\ &&   +2 \sqrt{2} f_1 v_\rho v_{s_2} (v_\eta^2+v_\chi^2)  
\end{eqnarray}

In this case the Goldstone boson is given by

\begin{equation}
G_2^-=\frac{1}{\sqrt{v_\chi^2+v_\eta^2+v_{s_2}^2}}(v_\chi, \, -v_\eta, \, v_{s_2})^T
\end{equation}

\subsection{Double charge scalars}
\label{subsec:2charges}

The mass matrix is $ M^2_{++}$, with the basis $(\chi^{++}\,\rho^{++}\,S^{++}_1\,S_2^{++})$ given by
\begin{eqnarray}
&& (M^2_{++})_{11}=\frac{v_\rho \left(-\sqrt{2} f_1 v_\eta-f_2 v_{s_2}+a_8 v_\rho v_\chi\right)}{2 v_\chi}, \nonumber \\ && 
(M^2_{++})_{12}=\frac{1}{2} \left(-\sqrt{2} f_1 v_\eta+f_2 v_{s_2}+a_8 v_\rho v_\chi\right), 
\nonumber \\ &&
(M^2_{++})_{13}= \frac{2 f_2 v_\rho+d_2 v_{s_2} v_\chi}{2 \sqrt{2}},
\nonumber \\&& 
(M^2_{++})_{14}= \frac{d_2 v_{s_2} v_\chi}{2 \sqrt{2}}, \nonumber \\ &&
(M^2_{++})_{22} =\frac{v_\chi \left(-\sqrt{2} f_1 v_\eta-f_2 v_{s_2}+a_8 v_\rho v_\chi\right)}{2 v_\rho}, 
\nonumber \\ && 
(M^2_{++})_{23}=\frac{d_6 v_\rho v_{s_2}}{2 \sqrt{2}},
\nonumber \\ && 
(M^2_{++})_{24}=\frac{d_6 v_\rho v_{s_2}+2 f_2 v_\chi}{2 \sqrt{2}}, \nonumber \\ && 
(M^2_{++})_{33}=\frac{2 e_2 v_{s_2}^3+d_6 v_\rho^2 v_{s_2}-v_\chi (2 f_2 v_\rho+d_2 v_{s_2} v_\chi)}{4 v_{s_2}},\nonumber \\ &&
(M^2_{++})_{34}=\frac{e_2 v_{s_2}^2}{2},  \nonumber \\ &&
(M^2_{++})_{44}= \frac{2 e_2 v_{s_2}^3-d_6 v_\rho^2 v_{s_2}+v_\chi (d_2 v_{s_2} v_\chi-2 f_2 v_\rho)}{4 v_{s_2}}, 
\end{eqnarray}

This mass matrix has only one analytical eigenvalue: $M^2_{++1}=0$. The respective
Goldstone boson is
\begin{equation}
G^{--}=\frac{1}{\sqrt{v_\rho^2+v_\chi^2+4v_{s_2}^2}}(v_\chi, \, -v_\eta, \, v_{s_2},\, -\sqrt{2}v_{s_2})^T
\end{equation}

	

\end{document}